\documentclass[pdftex,twocolumn,10pt,letterpaper]{extarticle}

\pdfoutput=1 %for arxiv
\newcommand{\AUTHORS}{Hafiz Mohsin Bashir\textsuperscript{1}, 
                    Abdullah Bin Faisal\textsuperscript{1}, 
                    Muhammad Asim Jamshed\textsuperscript{2}, 
                    Peter Vondras\textsuperscript{1}, 
                    Ali Musa Iftikhar\textsuperscript{1}, 
                    Ihsan Ayyub Qazi\textsuperscript{3}, 
                    Fahad R. Dogar\textsuperscript{1}
                    }

\newcommand{\TITLE}{Reducing Tail Latency via Safe and Simple Duplication}

\newcommand{\KEYWORDS}{Put your keywords here}
\newcommand{\CONFERENCE}{Somewhere}
\newcommand{\PAGENUMBERS}{yes}       % "yes" or "no"
\newcommand{\COLOR}{yes}
\newcommand{\showComments}{no}
\newcommand{\comment}[1]{}

\newfont{\ttlfnt}{phvb8t at 18pt}  %arxiv
\newfont{\aufnt}{phvr8t at 12pt}    %arxiv
\newfont{\auit}{phvro8t at 12pt}    %arxiv % GM 2/4/2000
\newfont{\affaddr}{phvr8t at 10pt}  %arxiv

\usepackage{tikz}
\newcommand*\circled[1]{\tikz[baseline=(char.base)]{%
            \node[shape=circle,fill=blue!20,draw,inner sep=2pt] (char) {#1};}}
\newcommand*\circledl[1]{\tikz[baseline=(char.base)]{%
            \node[shape=circle,fill=orange!20,draw,inner sep=2pt] (char) {#1};}}

\usepackage{listings}

\usepackage[T1]{fontenc}
\usepackage[utf8]{inputenc}
\usepackage{verbatim}
\usepackage{newtxtext,newtxmath}       % Times/Times-like math symbols
\usepackage{bm}                        % bold math; use \bm{} in captions
\usepackage[scaled=0.92]{helvet}
\usepackage{courier}
\usepackage{subfigure}
\usepackage[ruled,noline,noend]{algorithm2e}
\usepackage{supertabular}
\usepackage{tabularx}
\usepackage{wrapfig}
\usepackage{ifthen}
\ifthenelse{\equal{\PAGENUMBERS}{yes}}{%
\usepackage[nohead,
            left=1.0in,right=1.0in,top=1.0in,
            footskip=0.5in,bottom=1in,     % Room for page numbers
            columnsep=0.33in
		]{geometry}
}{%
\usepackage[noheadfoot,left=0.75in,right=0.85in,top=0.75in,
            footskip=0.5in,bottom=1in,
            columnsep=0.25in
	    ]{geometry}
}

\usepackage[font=bf]{caption}
\usepackage{titlesec}
\titlespacing{\paragraph}{0pt}{*1}{*1}      % SPACE
\titleformat{\paragraph}[runin]{\normalfont\normalsize\bfseries}{\theparagraph}{1em}{}

\titleformat{\section}%                     % ACM: caps + period (for 10pt doc)
  {\bf\large\uppercase}{\thesection.\quad}{0pt}{}
\titleformat{\subsubsection}%                % ACM
  {\it\large}{\thesubsubsection\quad}{0pt}{}
\usepackage{enumitem}
\setlist{itemsep=0pt,parsep=0pt}             % more compact lists

\usepackage{fancyhdr}

\ifthenelse{\equal{\PAGENUMBERS}{yes}}{%
  \pagestyle{plain}
}{%
  \pagestyle{empty}
}

\usepackage[numbers]{natbib}

\usepackage{booktabs}
\usepackage{multirow}
\usepackage{color}
\usepackage{colortbl}
\usepackage{float}                           % Must appear before hyperref to
\ifthenelse{\equal{\COLOR}{yes}}
{%
  \usepackage[colorlinks]{hyperref}%         % for online version
}
{%
  \usepackage[pdfborder={0 0 0}]{hyperref}%  % for paper (B&W) version
}
\usepackage{url}

\hypersetup{%
pdfauthor = {\AUTHORS},
pdftitle = {\TITLE},
pdfsubject = {\CONFERENCE},
pdfkeywords = {\KEYWORDS},
bookmarksopen = {true}
}

\usepackage{hyperref}
\hypersetup{
     colorlinks   = true,
     citecolor    = blue,
     urlcolor = blue,
}

\definecolor{placeholderbg}{rgb}{0.85,0.85,0.85}

\usepackage{graphicx}
\usepackage{soul} 
\soulregister\cite7
\soulregister\ref7
\soulregister\pageref7

\usepackage[absolute]{textpos}

\newfloat{acmcr}{b}{acmcr}

\newcommand{\note}[2]{
    \ifthenelse{\equal{\showComments}{yes}}{\textcolor{#1}{#2}}{}
}

\newcommand{\stage}{\textsf{D-Stage}}
\newcommand{\stages}{\textsf{D-Stages}}
\newcommand{\sched}{\textsf{DAS}}
\newcommand{\hkust}{\textsf{DP-Network}}
\newcommand{\snort}{\textsf{Snort}}
\newcommand{\etal}{\textit{et al.}}

\date{}
\title{\ttlfnt \TITLE }
\author{{\aufnt Hafiz Mohsin Bashir\textsuperscript{1}, 
                Abdullah Bin Faisal\textsuperscript{1}, 
                Muhammad Asim Jamshed\textsuperscript{2}, 
                Peter Vondras\textsuperscript{1},} \\
        {\aufnt Ali Musa Iftikhar\textsuperscript{1}, 
                Ihsan Ayyub Qazi\textsuperscript{3}, 
                Fahad R. Dogar\textsuperscript{1} 
                \vspace{1ex}}\\
        {\affaddr \textsuperscript{1}Tufts University, 
                    \textsuperscript{2}Intel Labs, 
                    \textsuperscript{3}LUMS
        }
       }
\begin{document}

\maketitle

%\AcmCopyright
%\ToAppear

\abstract{\vspace{-0.1in}
Duplication can be a powerful strategy for overcoming stragglers in cloud services, but is often used conservatively because of the risk of overloading the system. 
We present duplicate-aware scheduling or \sched{},
which makes duplication safe and easy to use, by leveraging the two well-known primitives of prioritization and purging. 
To support \sched{} across diverse layers of a cloud system (e.g., network, storage, etc), we propose
the \stage{} abstraction, which decouples the duplication policy 
from the mechanism, and facilitates working with legacy layers of a system. 
Using this abstraction, we evaluate the benefits of \sched{} for two data parallel applications (HDFS, an in-memory workload generator) and a network function (\snort{}-based IDS cluster). Our experiments on the public cloud and Emulab show that \sched{} is safe to use, and the tail latency improvement holds across a wide range of workloads.

}
\section{Introduction}
\label{sec:intro}

Meeting the performance expectations of cloud applications is challenging: typical cloud applications have workflows with high fanout, involving multiple potential bottleneck resources (e.g., network, storage, processing, etc), so even a single slow component, or \emph{straggler}, ends up delaying the entire application. 
Studies show that stragglers can cause significant increase in tail latency~\cite{dctcp12, baraat, dolly, hopper}.
Because a slew of unpredictable factors (e.g., failures, load spikes, background processes, etc) can cause stragglers~\cite{tail_at_scale, dolly, bobtail}, overcoming them is difficult, especially for application workloads with small requests that only span tens to hundreds of milliseconds -- for such applications, the time to detect and react to stragglers is typically too short~\cite{mittos}.

A powerful approach for dealing with stragglers is to avoid them using \emph{duplication}. 
Duplication can leverage the inherent redundancy present at different levels of a typical cloud system -- redundant network paths in a data center topology or replicated
application and storage servers -- to overcome a broad range of straggler scenarios.   
For example, Dean et al. 
describe a collection of ``tail tolerant'' techniques used at Google, which duplicate 
\texttt{get()} requests for large scale data parallel applications~\cite{tail_at_scale}.  
At the \emph{network} level, prior work has shown the benefits of duplicating 
flows (or specific packets of a flow)~\cite{repflow, lowlatency, repnet, rans-hotnets}. 
Similarly, other systems have shown the efficacy of duplication for \emph{storage} (e.g.,~\cite{mittos, scads, zoolander}) and \emph{distributed job execution frameworks}~\cite{late, mantri, pbse, dolly}.

Despite the potential benefits of duplication,
its use is fraught with danger:
the extra load caused by duplication can degrade system performance
or even make the system unstable. 
For example, a duplicate \texttt{get()} request will not just create extra work for the application and storage servers, but also increase load on other resources (e.g., network, load-balancer, etc). Because of this danger, existing systems typically use duplication in a conservative manner, employing techniques that \emph{selectively} issue duplicates. 
For example, in \textsf{hedged-request}~\cite{tail_at_scale}, a duplicate is issued only if the primary does not finish within a certain time (e.g., $95^{th}$ percentile expected latency).
Such heuristics can turn out to be adhoc, as workload and system load changes, making today's multi-layered cloud systems even more brittle and complex.

\begin{comment}
For example, a duplicate get() request may not just overload
the storage or processing servers, but can also cause increased contention on other (shared) resources (e.g., network or metadata server).    
Similarly, the threshold to issue a duplicate 
get() request (e.g., 95$^{th}$ percentile) is highly dependent on the system load and workload, which may be highly variable.
\end{comment}

We argue that these challenges stem from lack of explicit support for duplication, and call for making duplication a first-class concept in modern cloud systems: it should be easy to specify duplication requirements, 
and different layers of the system should have explicit support for meeting these requirements. Toward this goal, we identify three important questions that need to be answered first: i) can we have duplication \emph{policies} that are safe and easy to use, i.e., they don't require careful tuning of thresholds, nor do they overload the system, ii) can we design \emph{abstractions} that make it easy to support diverse duplication policies at different layers of the system? and iii) how can we effectively deal with \emph{legacy layers} that cannot be modified to support duplication? Our work seeks to answer these questions and makes two contributions. 

Our first contribution is a new duplication policy, duplicate-aware scheduling or \sched{}, which combines \emph{prioritization} and \emph{purging} to make duplication easy and safe to use. Prioritization ensures that duplicates are treated at a lower priority and don't harm primaries while purging ensures that unnecessary copies in the system are removed in a timely manner so they do not overwhelm any auxiliary resource in the system.  In this way, \sched{} is grounded in theory and informed by practice:  it leverages key recent results in scheduling theory~\cite{gardnerthesis}, which shows the benefits of prioritization when duplicating requests are present in the system, with the insights from practical, large-scale systems, which show the need (and benefits) of purging~\cite{tail_at_scale}.

Our second contribution is an abstraction, duplicate-aware stage (\stage{}), that makes it easy to support \sched{} at different layers of a cloud system. 
A \stage{} comprises of \emph{queues} with suitable duplication controls that are necessary
for supporting \sched{} (and other duplication policies). 
The \stage{} abstraction provides three key benefits. 

First, there is a \emph{decoupling} between the duplication policy and mechanism, through a high level interface that exposes key duplication policy controls -- such as the number of duplicates, their priorities, when and how they should be dispatched and purged, etc -- while hiding the details of their implementation mechanism, such as what prioritization or purging mechanism is used.

Second, a \stage{} operates on the notion of a \emph{job}, which has an associated \emph{metadata} that consistently identifies the duplication requirements of a job as it travels across different layers of the system -- e.g., from an application request to network packets and I/O requests --  enabling consistent treatment of duplicates, such as differentiating between primary and duplicates or not duplicating an already duplicate job. 

Third, a Proxy \stage{}, which is a special type of \stage{}, helps in dealing with the challenge of legacy layers that may not be amenable to modification. A Proxy \stage{} is inserted in front of a legacy layer, in order to approximate the duplication functionality.  It has all the functionality of a typical \stage{} but also supports \emph{throttling} of jobs going into the legacy layer, keeping as many jobs in its own queues, in order to retain control over key duplication primitives, such as prioritization and purging.

We validate our contributions in the context of challenging workloads involving data-parallel applications and networks functions (NF). For data-parallel applications, we use the Hadoop Distributed File System (HDFS) and \hkust{}~\cite{hkust-tg-paper}, a research prototype used for small in-memory workloads (e.g., web search~\cite{dctcp12}). 
For both applications, a \texttt{read()/get()} request is duplicated and each copy is sent to one of the multiple available replicas; in this scenario,  data could be served from disk or memory, which could create different bottlenecks in the system (e.g., network, storage). We implement \stages{} support for storage and network layers, leveraging existing mechanisms for prioritization and queuing that are available in commodity servers (e.g., CFQ disk scheduler~\cite{kerneldoc-CFQ}, Priority Queues, etc).
For NF evaluation, we consider a distributed IDS cluster scenario, where CPU is the main bottleneck, and \stages{} at the network and processing levels are required to get the full benefits of duplication. 

We evaluate these applications using various micro and macro-benchmark experiments on Emulab~\cite{emulab} and the Google Cloud~\cite{gcloud}.
Our experiments on Google Cloud validate the presence of stragglers, and \sched{}'s ability to overcome them. We also find that \sched{} is seamlessly able to avoid hotspots, which are common in practical workloads, by leveraging an alternate replica, thereby obviating the need for sophisticed  replica selection techniques~\cite{c3}. Using controlled experiments on Emulab, we show that \sched{} is safe to use: even under high loads, when many existing duplicate schemes make the system unstable, \sched{} remains stable and provides performance improvements. Our results across a wide range of scenarios show that \sched{}'s performance is comparable to, or better than, the best performing duplication scheme for that particular scenario. 

In the next sections, we describe our contributions\footnote{Our work does not raise any ethical issues.} and how we validate them. 
Overall, we view our work as taking an important step towards making duplication a first-class concept for cloud systems. As we discuss in \S\ref{sec:discuss}, our work also opens up several interesting directions for future work, such as implementing the \stage{} abstraction for other resources (e.g., end-host network stack) and potentially combining the work done by the primaries and duplicates to improve system throughput. 

\section{Duplicate-Aware Scheduling}
\label{sec:scheduling}

\begin{table*}
\centering
\resizebox{\textwidth}{!}{%

\begin{tabular}{|c|c|c|l|}
\hline
\textbf{} & \textbf{Duplication Techniques} & \textbf{Example Schemes} & \multicolumn{1}{c|}{\textbf{Comment}} \\ \hline
Full Duplication & Cloning & SCADS~\cite{scads}, CosTLO~\cite{costlo} & Simple to implement but doubles the system load \\ \hline
\multirow{3}{*}{\begin{tabular}[c]{@{}c@{}}Selective \\ Duplication\end{tabular}} & Speculative Duplication & \begin{tabular}[c]{@{}c@{}}LATE~\cite{late},  Mantri~\cite{mantri}\\ PBSE~\cite{pbse}, Hedged/AppTO~\cite{tail_at_scale}\end{tabular} & 
\begin{tabular}[c]{@{}l@{}} $\blacktriangleright$ Wait before speculation limits the ability to deal with stragglers in small jobs\\ $\blacktriangleright$ Requires sophisticated speculation mechanism
\end{tabular} \\ \cline{2-4} 
 & Small Job Only & Dolly~\cite{dolly}, RepFlow~\cite{repflow}, RepNet~\cite{repnet}& \begin{tabular}[c]{@{}l@{}} $\blacktriangleright$ Only effective for small jobs\\ $\blacktriangleright$ Classifying a job as small apriori may not be possible for some workloads
 \end{tabular} \\ \cline{2-4} 
 & Shorter Queue Selection & MittOS~\cite{mittos}, Tied~\cite{tail_at_scale} & Can only handle queue overloads and not other stragglers (e.g., noise~\cite{bobtail}). \\ \hline
 
\end{tabular}}
\caption{Comparison of various duplication schemes. Existing schemes tend to be either simple or safe, but not both.}
\label{table-dup-schemes}
\end{table*}

\subsection{Existing Duplication Schemes}
\label{subsec:existing-schemes}

A large body of recent work focuses on using duplication for straggler mitigation in cloud systems. 
We differentiate our work from these proposals on three fronts:

\noindent
\circled{1} \textbf{Duplication Across Multiple Resources.}
Existing duplication proposals focus on a particular bottleneck resource (layer) (e.g., storage, network, etc)~\cite{mittos, tail_at_scale, repflow, repnet}. To the best of our knowledge, we are the first to strive for making all important layers of the cloud system duplicate-aware, i.e., they need to handle duplicates in a coordinated manner. 

\noindent
\circled{2} \textbf{Millisecond-level Workloads.}
Several existing proposals (e.g., Dolly~\cite{dolly}, LATE~\cite{late}, etc) focus on workloads where jobs last for seconds or even longer.
In contrast, \sched{}'s scope includes millisecond-level workloads -- as argued in recent work~\cite{mittos}, such workloads are more challenging and require a proactive approach to duplication.

\noindent
\circled{3} \textbf{Safe and Easy Duplication.}
Table~\ref{table-dup-schemes} divides existing proposals into two broad categories based on how they duplicate jobs: i) In \emph{full-duplication} schemes, every job is duplicated, so they are easy to use, but dealing with the extra load of the duplicates and keeping the system stable under high load is a major challenge. ii) In \emph{Selective Duplication} schemes (e.g., Dolly~\cite{dolly}, Tied~\cite{tail_at_scale}, etc) only a fraction of requests are duplicated (based on some criteria). For example, in the case of \textsf{Hedged Request}~\cite{tail_at_scale}, a duplicate of a job is launched if the job fails to finish within the 95$^{th}$ percentile of its expected latency. Such techniques are usually safe, but can be complex, typically requiring careful tuning of thresholds or resource specific optimizations. 

To validate \circled{3}, 
we conduct a simple experiment on a small scale HDFS cluster on Emulab~\cite{emulab}.
A client retrieves objects of 10MB size from this cluster under the following schemes: \textsf{Single-Copy} (no duplication), \textsf{Cloning} (full duplication), \textsf{Hedged-Low} (optimized for low load), and \textsf{Hedged-High} (optimized for high load). To create stragglers, we used the noise model discussed in details in~(\S\ref{sec:eval})).
%To create stragglers, we induced ``noise'' for roughly 10\% of the requests (details in~(\S\ref{sec:eval})). 

\begin{figure}[!t]
  \begin{center}
    \includegraphics[width=0.40\textwidth]{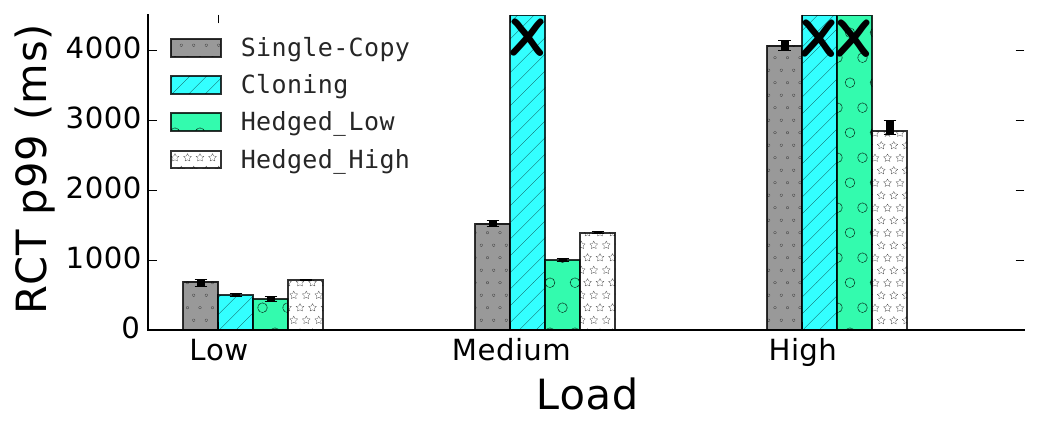}
  \end{center}
   \vspace{-0.2in}
  \caption{A simple experiment to validate the key findings of Table~\ref{table-dup-schemes}. Max-sized bar with black cross indicates that the corresponding scheme becomes unstable.}
  \label{fig:das-motivation}
\end{figure}

Figure~\ref{fig:das-motivation} shows that at low load
\textsf{Cloning} effectively reduces the 99$^{th}$ percentile of request completion times (RCT), but it becomes unstable at medium and high loads, i.e., the queues keep growing and requests don't finish. 
In contrast, the selective duplication schemes only work well for the load they are optimized for but can perform poorly otherwise. For example,  \textsf{Hedged-Low} provides benefits at low load while \textsf{Hedged-High} performs well at high load. However, at high load \textsf{Hedged-Low} becomes unstable as it issues too many duplicates; it highlights that the threshold should be highly optimized for a particular scenario, which can be challenging to get right for realistic scenarios. In summary, schemes like \textsf{Cloning} are simple but not safe, while selective schemes like \textsf{Hedged} are (usually) safe but difficult to get right. 

\subsection{\sched{}}
\label{subsection:model}
We identify two requirements for an ideal duplication policy:
\noindent
\circledl{1} \textbf{Simple.} Similar to full-duplication schemes, the decision of when to issue a duplicate and how many duplicates to issue should be simple: it should not require careful selection of thresholds, as in selective duplication schemes. 

\noindent    
\circledl{2} \textbf{Safe.} Its stability properties should be similar to a system that does not use duplication: issuing duplicates should not harm the system under any situation -- of course, while reaping the benefits of duplication whenever possible.  

We propose duplicate-aware scheduling (\sched{}), a simple and safe duplication policy. 
It is simple because each (primary) job is duplicated once, typically as it enters the system, 
and then each layer provides prioritization and purging to make the system safe. 

\paragraph{Prioritization.} Primary requests get \emph{strict priority} over duplicates,
ensuring preemption and work-conservation, i.e. duplicate jobs are served if and only if there is no primary in the system, and the duplicate
 is preempted when a primary arrives. This behavior ensures that no duplicate delays a primary: the key insight is that prioritization between primaries and duplicates provides the ability to seamlessly adapt to variations in system load i.e., under low load, duplicates have a higher chance of completion, which improves response times, whereas under high overload, primaries are shielded from duplicates.

While the use of prioritization is quite common in the networking community~\cite{lowlatency, repflow, pase}, 
recent advances in scheduling theory \cite{GARDNER2017,gardnerthesis} show that a \sched{}-like duplication system, called Primaries First (PF), which uses prioritization can achieve two strong properties.

\begin{enumerate}[leftmargin=*,noitemsep,topsep=0ex]
    \item \textit{Safety and Performance.} It performs better than a \textsf{Single-Copy} scheme -- every job achieves a response time at least as low as under a system without duplication (\textsf{Single-Copy}\footnote{We assume jobs are served in FIFO order.}). This property holds for correlated service times and any job size distribution~\cite{gardnerthesis}. Further, under simplistic assumptions (i.e., exponentially distributed job sizes and independence across servers), it can achieve near-optimal mean response time~\cite{GARDNER2017}. 
    Our evaluation shows that \sched{} achieves low response times even when these assumptions are relaxed (e.g., under correlated service times and heavy-tailed job size distributions).

\smallskip

    \item \textit{Fairness.} Another desirable feature is that this policy is fair to non-duplicate jobs, i.e., those jobs that don't make use of duplication. 
    For duplicate-aware systems, we adopt the following notion of fairness from \cite{refair}: \emph{policy $\pi$ is fair if, under policy $\pi$, no job class experiences a higher mean response time than under a baseline policy $\eta$, where our baseline is} \textsf{Single-Copy}. Gardner et al. \cite{GARDNER2017} showed that PF satisfies this notion of fairness between duplicate and non-duplicate classes.
    
\smallskip

\end{enumerate}

\paragraph{Purging.} 
While scheduling theory may suggest that prioritization is sufficient to meet our requirements, for practical systems, we also need purging: jobs that are no longer required -- duplicate jobs whose primaries have finished (or vice versa) -- should be removed (purged) from the system. 
%Even with prioritization, it is essential to have purging because it ensures that any \emph{auxiliary} system resource does not become a bottleneck.
Purging is essential because it ensures that any \emph{auxiliary} system resource does not become a bottleneck.
For example, duplicate jobs waiting in a lower priority queue may cause exhaustion of transmission control blocks (TCBs), limit buffer space for primary jobs, or increase contention overhead of thread scheduling. Overall, purging improves performance and system stability, especially under high load, by eliminating unnecessary jobs from the system. 

While \sched{} promises to be both simple and safe, in order to achieve these benefits, we need to support \sched{} at \emph{every} potential bottleneck layer of the system; a seemingly daunting task made easier by the \stage{} abstraction, which we present next. 
\section{The \stage{} Abstraction}
\label{sec:design}

\begin{figure}[!t]
  \begin{center}
    \includegraphics[width=0.49\textwidth]{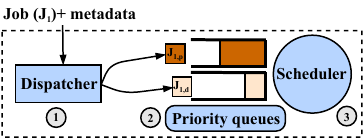}
      \end{center}
  \caption{The \stage{} abstraction has three components. Job + metadata enters the stage, the dispatcher (1) creates a duplicate copy (if required) and puts the job(s) in their respective priority queues (2). A scheduler (3) schedules them in a strict priority fashion. Jobs can also be purged (now shown).}
  \label{fig:abstraction}
\end{figure}

\begin{comment}

 \begin{table}[t]
\small
 \texttt{dispatch (job, metadata)} \\
  \texttt{schedule()} \\
  \texttt{purge (job-id(s), CASCADE\_FLAG)} \\
  \hline
 \caption{\stage{} Interface}
 \label{tab:interface}
\end{table}
\end{comment}

\begin{table}
\begin{tabular}{l}
\texttt{dispatch (job, metadata)} \\
\texttt{schedule()} \\
\texttt{purge (job-id(s), CASCADE\_FLAG)} \\ \hline
\end{tabular}
\caption{\stage{} Interface}
\label{tab:interface}
\end{table}

A \stage{} comprises of \emph{queues}, which are inserted at potential bottleneck layers of a system, providing the necessary \emph{control} for duplication, such as when to create a duplicate, its priority, and support for purging. 
Figure~\ref{fig:abstraction} zooms into a \stage{}: 
each \stage{} operates on a job with associated metadata (e.g., id, priority) that consistently identifies the job across different layers of the stack, and supports key duplication controls through a high level interface (Table~\ref{tab:interface}).
The \stage{} abstraction provides three key benefits:

\paragraph{1. Making duplication visible across layers.} 
While the notion of a job changes across layers -- such as a flow for a transport \stage{} and a packet for a network \stage{} -- the associated metadata with a job (e.g., its id, priority etc) consistently identifies that job and its duplication requirements. This ensures that a job can be tracked across layers (e.g., for purging), gets differential treatment (e.g., lower priority for duplicate jobs), and layers don't end up making unnecessary duplicates of jobs that have already been duplicated. This simple requirement of associating the right metadata with jobs is crucial for making duplicates a first class concept in cloud systems.

\paragraph{2. Decoupling of policy and mechanism.} 
By identifying the key duplication primitives (dispatching, scheduling, and purging) and providing control over them through a high level interface, a \stage{} decouples the duplication policy from the mechanism. As a concrete example, the policy may specify that duplicates should be given strictly lower priority, but the mechanism will decide the best way to support such prioritization. 
The benefits of this separation of concerns is well known in systems design -- it is even more crucial for duplication in today's cloud systems which involve many diverse layers: it will be impractical for system administrators, who want to specify policies, to implement duplication mechanisms that are optimized for specific resources, such as storage, network, etc.   

\paragraph{3. Support for legacy layers.} 
Introducing \stage{} into existing layers could be challenging: modifying a layer may be infeasible, and some layers may not be amenable to purging or queuing of jobs (e.g., MongoDB~\cite{mittos}). 
A \stage{} abstraction can also support a Proxy \stage{}, which can approximate the behavior of a legacy \stage{}. The key idea is to throttle jobs going into the legacy layer to retain control over prioritization and purging in a preceding Proxy \stage{}. %(\S\ref{subsec:proxy})

\subsection{End-to-End Use Cases}
\label{subsec:target-scen}

For a typical cloud application, a \stage{} would communicate with other \stages{}, forming a ``network'' of \stages{}. 
We describe two diverse use-cases for deploying \stages{}, which we focus on in this paper:

\paragraph{1- Data Parallel Applications.}

\begin{figure}[!t]
  \begin{center}
    \includegraphics[width=0.49\textwidth]{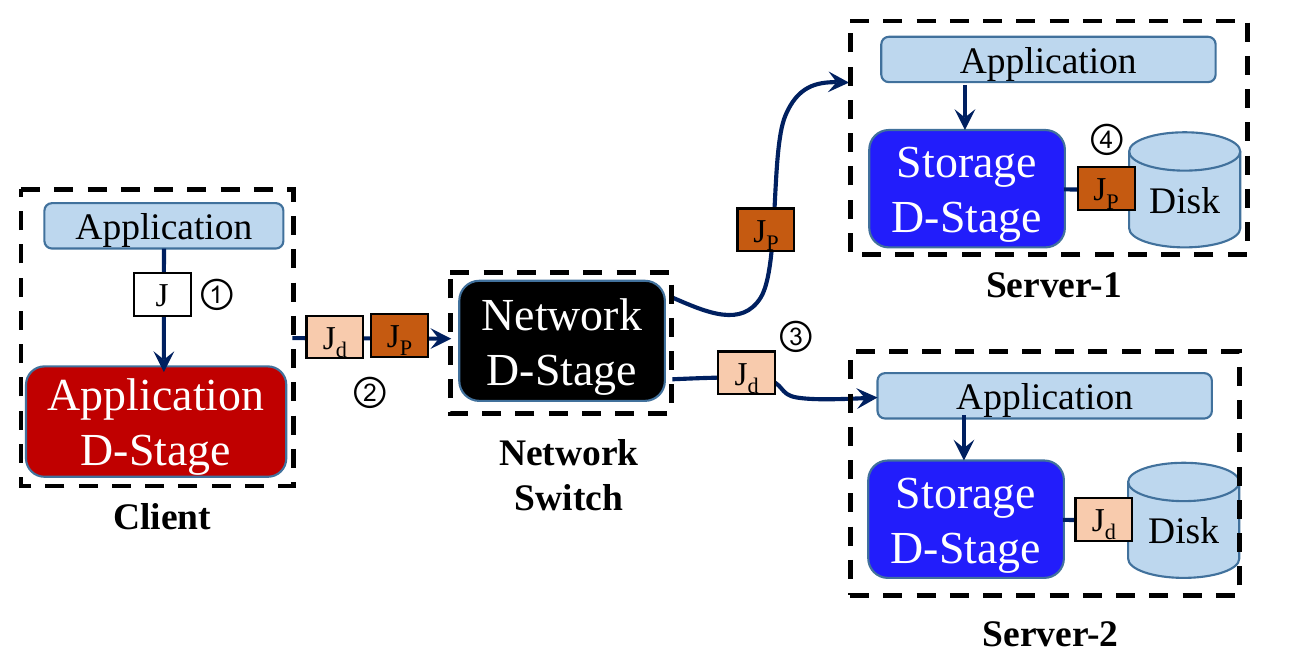}
      \end{center}
%   \vspace{-0.1in}
  \caption{Example showing how multiple \stages{} can be combined to serve a get() request from a client application.}
  \label{fig:getfile}
\end{figure}

Figure~\ref{fig:getfile} shows an example for a \emph{get()} operation in data parallel applications (e.g., MongoDB, HDFS). An application layer \stage{} at the client gets the request, it creates a duplicate request and sends both the primary and the duplicate to a subsequent \stage{}, scheduling them based on their priority. Other \stages{} in this example  -- like the network and storage -- don't create any more duplicates: they just provide differential treatment (prioritization) to primaries and duplicates. Once the primary (or duplicate) finishes, the client \stage{} purges the other copy, with all subsequent \stages{} removing any corresponding job(s) from their queues.

\begin{figure}[!t]
  \begin{center}
    \includegraphics[width=0.49\textwidth]{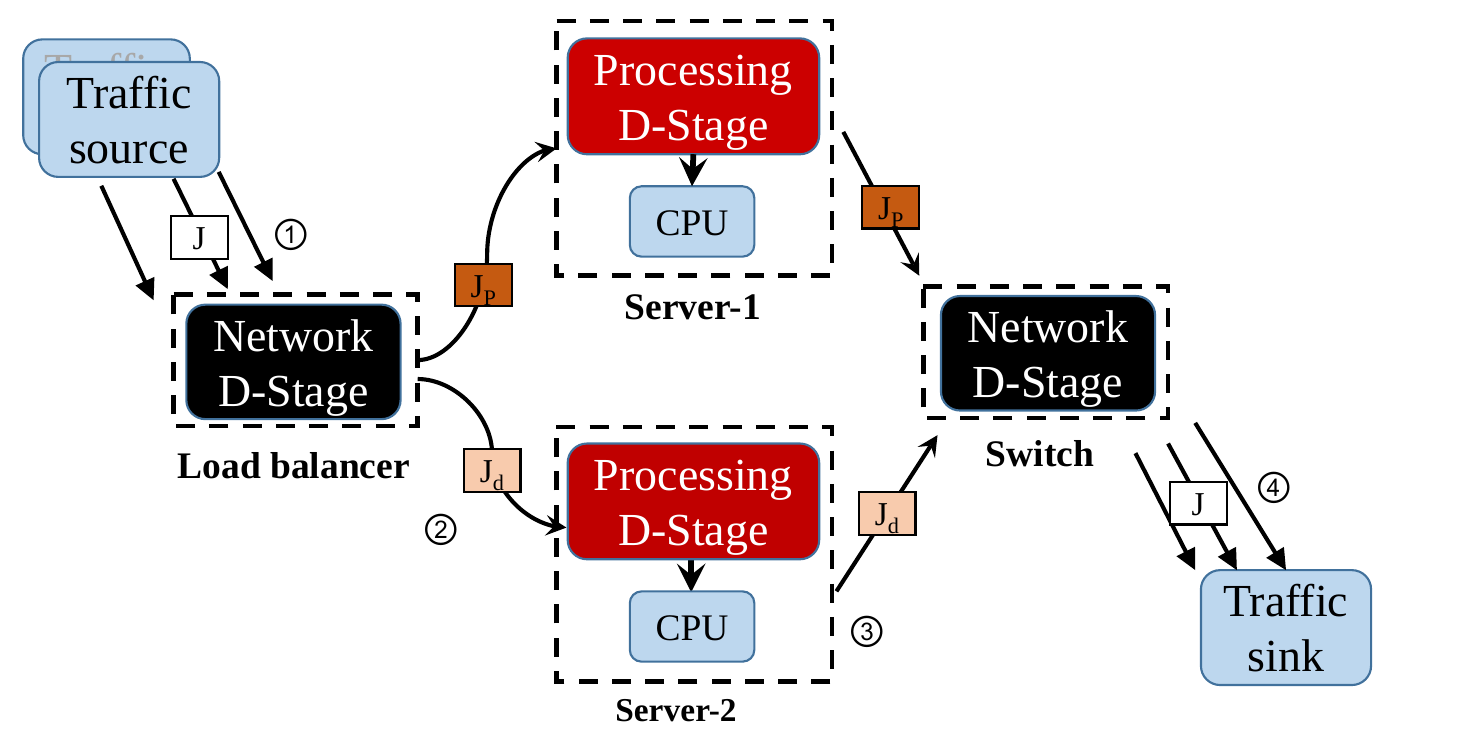}
      \end{center}
%   \vspace{-0.1in}
  \caption{Example showing how \stages{} can be placed at different layers for supporting NFs (e.g., an IDS cluster).}
  \label{fig:proc_pkts}
\end{figure}

\paragraph{2- Network Function Virtualization (NFV).}

NFV deployments are common in today's cloud environments. For example, an IDS cluster in the cloud consists of multiple nodes responsible for detecting malicious traffic. For this, IDS nodes are required to pass incoming packets through complex state machines and thus they are CPU bound.

Figure~\ref{fig:proc_pkts} shows how \stages{} can be used in such scenarios. A network \stage{} (e.g., load balancer) on the critical path of packet streams creates duplicate flows and forwards the primary and duplicate flows to different nodes in the NFV cluster. 
A processing \stage{} running at each NF node processes packets in a priority-aware fashion: duplicates are only processed if there is spare capacity on that node. Yet another network \stage{} is responsible for taking processed packets, and filtering out duplicates before forwarding them to the next stage, which could be a cluster of web-servers running a data-parallel application.  

\vspace{0.1in}
In the following subsections, we elaborate on the key features of a \stage{} which enable duplication benefits, followed by case studies on how we setup \stages{} to enable \sched{} support in three different applications (\S\ref{subsec:case-studies}).

\begin{comment}
\begin{table}[thbp]
\begin{tabular}{l}
\small
\texttt{job-id //identifies the job}\\
\texttt{priority //priority of the job} \\
\texttt{start-time //arrival time in the system} \\
\texttt{sched-time //wait time before scheduling} \\
\texttt{duplication() //creating duplicate copies} \\
\texttt{atStart() //processing when its turn comes} \\
\texttt{atEnd() //processing when done} \\ \hline
 \end{tabular}
 \caption{Job Metadata}
 \label{tab:metadata}
 \end{table}
\end{comment}

\begin{table}[thbp]
\begin{tabularx}{\columnwidth}{l}
\small
\texttt{job-id //identifies the job}\\
\texttt{priority //priority of the job} \\
\texttt{start-time //arrival time in the system} \\
\texttt{sched-time //wait time before scheduling} \\
\texttt{duplication() //creating duplicate copies} \\
\texttt{atStart() //processing when its turn comes} \\
\texttt{atEnd() //processing when done} \\ \hline
 \end{tabularx}
 \caption{Job Metadata}
 \label{tab:metadata}
 \end{table}

\subsection{Job and Metadata}
\label{subsec:job}
As noted earlier, each layer has its own notion of a job. For example, at the network \stage{}, the job could be a packet while an application layer \stage{} may operate on \emph{read{}} requests for files/objects. Each job should have an associated \emph{metadata} that contains 
duplication specific information:

Table~\ref{tab:metadata} lists the metadata required by a \stage{}. 
A \emph{job-id} is used to identify a job -- it should be unique and consistent across \stages{}, so a \stage{} can purge a job which is enqueued inside another \stage{} using the \emph{job-id}. 
To work with legacy layers, a \stage{} can also maintain mapping of its own job-ids to the job-ids used by the legacy layer. For example, an application layer \stage{}
can maintain the mapping of application-level request 
to its corresponding flow or socket identifier that is used at the transport layer. 

The \emph{priority} of a job is used to decide how this job will be scheduled. Many jobs already have a notion of priority, even if it is rarely used. For example, the ToS bit in the packet header can be used to determine the priority of a packet. 
The \emph{start-time} and \emph{sched-time} are useful in making scheduling and purging decisions. For example, a scheme may decide to purge jobs which have been outstanding for a certain time; similarly, a scheme
like \textsf{hedged-request} may want to delay the scheduling of a duplicate until a certain time.

Finally, the metadata includes three functions (callbacks): i) \emph{duplication()} function, which contains the duplication logic, 
%such as whether the job should be duplicated, 
such as whether to duplicate the job,
and if yes, the number of duplicates to create, and any metadata required for each duplicate copy, such as its priority, new name, etc, ii) \emph{atStart()} function, which implements any logic that needs to be executed when a job is \emph{scheduled} (e.g., purge corresponding copy, as in \textsf{Tied-Request}) and iii) \emph{atEnd()} function, which provides similar support for any action that needs to be taken once the job is \emph{finished} executing; again this can be useful for purging the corresponding copy of the job.
 
\subsection{Interface}
\label{subsec:interface}
We identify key duplication primitives that can be combined to support a wide range of duplication policies. 
These primitives are used through a high-level interface (Table~\ref{tab:interface}), hiding the implementation details from other system layers. We describe the interface, and comment on its use for common types of \stages{}, such as the network, storage, and processing. 

\medskip
\noindent
\textbf{\texttt{dispatch(job, metadata)}}.
The dispatcher controls how jobs are placed inside the \stage{} queues. The dispatcher interprets and acts on the \emph{metadata}: it creates the necessary duplicate copies using the \emph{duplication()} callback, and puts both the primary and duplicate jobs in their respective queues.
For example, in Figure~\ref{fig:abstraction} the job $J_{1}$ is duplicated as job $J_{1,p}$ with high priority and $J_{1,d}$ with low priority -- $J_{1,p}$ goes into the high priority queue and $J_{1,d}$ is enqueued in low priority queue. Jobs with a later \emph{start-time} are put in a special delay queue (with an associated timer), where they wait until the timer expires or are purged from the queue. 

For some stages, where the duplication information cannot be embedded within the metadata, the dispatcher can have its own configurable rules for job duplication. For example, a dispatcher for a network \stage{} may use a hash of the header fields in a match-action fashion to create and dispatch duplicate copies to their suitable destinations.

In general, we do not expect the dispatcher to become a bottleneck for a \stage{}. However, under high load it can stop duplicating jobs, following the \sched{} principle that duplicates are strictly lower priority. 
Finally, as we describe in \S\ref{subsec:case-studies}, most practical scenarios would involve only one \stage{} duplicating the job in the end-to-end path while other \stages{} will simply operate on the primary and duplicate copies. 

\medskip
\noindent
\textbf{\texttt{schedule()}}. A \stage{} requires a \emph{priority scheduler} so as to provide differential treatment to primaries and duplicates based on their priority. A job's turn should be determined by its priority, and once it is scheduled, the scheduler should first execute \texttt{atStart()}, 
then process the job (just like it would do normally), 
and finally call the \texttt{atEnd()} function. An ideal priority scheduler should ensure strict priority -- with \emph{preemption} and \emph{work conservation} -- while incurring minimal overhead.
A desirable feature in this context is to have the ability to break a large job into smaller parts, so that the scheduler can efficiently preempt a lower priority job and then later resume working on it in a work conserving manner.
Fortunately, many layers already support some form of priority scheduling.  
For example, for storage applications, Linux supports I/O prioritization using \emph{completely fair queuing} (CFQ) scheduler. Similarly, for CPU prioritization, we can use support for thread prioritization on modern operating systems -- for example, our implementation uses the POSIX call $pthread\_setschedprio$~\cite{manpage-setschedprio}, which can prioritize threads at the CPU level.
Finally, modern switches and end-host network stacks already provide support for prioritization, where some header fields (e.g., ToS bit) can be used to decide the priority of flows and packets.

\medskip
\noindent
\textbf{\texttt{purge(job(s), CASCADE\_FLAG)}}. This interface supports specifying one or more jobs that should be purged by this \stage{}. 
An ideal purging implementation would allow purging of jobs from both the queues as well as the underlying system while they are being processed. The job(s) can be specified based on a predicate on any metadata information, such as matching a \emph{job-id} or those jobs that started before a certain \emph{start-time}. 
The CASCADE\_FLAG specifies whether the purge message should propagate to a subsequent stage, if the current stage is already done with processing the job and can no longer purge it.  
For example, an application may call a \emph{purge()} on the transport flow, which would result in the transport \stage{} purging its corresponding data from the end-host buffer, and if the flag is set, it will also call a \emph{purge} on the subsequent network \stage{}, so the packets that have left the end-host (and are inside the network) can be purged.

\subsection{Proxy \stage{}}
\label{subsec:proxy}

\begin{figure}[!t]
  \begin{center}
    \includegraphics[width=0.38\textwidth]{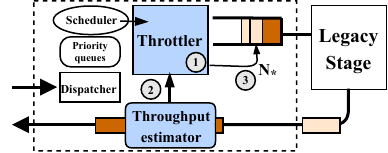}
      \end{center}
   \vspace{-0.1in}
  \caption{Proxy \stage{} sitting in front of a legacy stage. In addition to the components of traditional \stages{}, it has a throttler (1) which gets feedback from a throughput estimator (2) monitoring the output rate of the legacy stage. Based on this estimate, it discovers the multiplexing level (3).}
  \label{fig:feedback}
    \vspace{-0.1in}
\end{figure}

To support legacy layers in a system, we design a Proxy \stage{}, which is a specific implementation of a \stage{} and sits in front of a legacy (unmodified) stage; its goal is to retain maximal control over jobs going into the legacy stage by keeping them in its \emph{own} queues.
This allows the Proxy \stage{} to \emph{approximate} duplication support of an unmodified stage, using its own mechanisms for prioritization and purging. 

Figure~\ref{fig:feedback} shows the high-level working of a Proxy \stage{}.
Unlike a traditional \stage{}, it uses a special scheduler that \emph{throttles} outgoing jobs, based on the output rate of the legacy stage. Thus, the proxy \stage{} needs to be placed at a location where it can control and observe the input and output of the legacy stage.

\paragraph{Objective.} Choosing a suitable multiplexing level is important; a small value can lead to under-utilization of the unmodified stage whereas a large value would reduce queuing inside the proxy, thereby lowering its control (prioritizing, purging).
This trade-off crystallizes into the following objective: \emph{maximize} utilization (of the unmodified stage) while \emph{minimizing} the level of multiplexing (i.e., dispatching as few jobs as possible).

\paragraph{Throttling algorithm.}
Our desired objective closely matches with prior work on finding the right number of transactions to dispatch to a database~\cite{mpldb}.
In their work, Bianca et al. use offline queuing analysis to initialize the multiplexing level, which is then adjusted using feedback control in an online setting.

We take a similar approach but for simplicity the throttler is initialized with a multiplexing level of one.
We use the throughput of the legacy stage observed by the proxy to modulate the multiplexing level in the following way:

\begin{enumerate}[leftmargin=*,noitemsep,topsep=0ex]
\item \textbf{Probing phase.}
The throttler increases the multiplexing level as long as it observes commensurate increase in the throughput.
Otherwise, it stops and enters the \emph{exploitation} phase.

\item \textbf{Exploitation phase.}
Having discovered the ``best policy'', the throttler continues dispatching requests as per the optimal target\footnote{This is the minimum number of requests needed to maximize utilization.} ($N_{*}$).
\end{enumerate}

One of our primary use cases for the Proxy \stage{} is its use with an unmodified HDFS server (data-node), which has no notion of duplicates (\S\ref{sec:eval}). 
An HDFS cluster in a cloud setting could be bottle-necked anywhere.
Thus, we empirically verified that our throttling algorithm was able to determine the right multiplexing level for: i) SSD and HDD, if disk is the bottleneck, and ii) network links, if the workload is being served from cache.
\subsection{Case Study: Data Parallel Applications}
\label{subsec:case-studies}

We now describe how can we combine multiple \stages{} to implement \sched{} for two data parallel applications: i) HDFS~\cite{hdfs} and ii) an in-memory research prototype, which we refer to as \hkust{}~\cite{hkust-tg-paper}.
\hkust{} comprises of a client and server applications. The client application makes a request to transfer a specified flow size to the server, which sends the response, using a pool of persistent TCP connections between the client and the servers. \hkust{} has been 
used extensively in prior works (e.g., ClickNP~\cite{clicknp}, MQ-ECN~\cite{hkust-tg-paper}, FUSO~\cite{fuso}, etc) to evaluate performance of network protocols over small-scale (~40 nodes), 10G clusters. 

While both applications have similarities, there are also key differences in terms of how the \stages{} are supported within these applications (e.g., the purging mechanism and the use of Proxy \stage{}).
We first describe the common workflow of these (and other data parallel applications) and how we insert \stages{} at different layers. We then describe the specific support that we added for both of these applications. 

We divide the steps required to enable \sched{} support into two parts: the request phase, and the response phase.

\paragraph{Request Phase.} The first \stage{} is the \emph{request handler} stage, which is a processing \stage{} responsible for duplicating \emph{get()} request. The API between this stage and the application is the typical \emph{get()} API, which is enhanced to carry information about replicas and their priorities. As a processing \stage, it has support for thread prioritization, so a lower priority thread can work on sending the lower priority \emph{get()} requests, which ensures that the request generation part for duplicate requests does not hurt the primary requests. This is the \emph{only} stage in the \emph{get()} pipeline where duplication is involved. For all other stages, the dispatcher only places the job in its respective priority queue.

Requests are sent over the network \stage{} as per their priority until they reach their respective server node. On the server side, there is a request handler processing \stage{}, which is similar to the request handler stage of the client, except that it will not duplicate the job, as indicated by the metadata for the request. The request handler stage will then pass on the request to a storage \stage{}, which will follow the functionality described earlier. 

\paragraph{Response Phase.} The responses sent by the primary and secondary replicas traverse the network, using their appropriate priorities -- high priority for responses from the primary and low priority for responses from the secondary. On the client side, there is another processing \stage{} called the \emph{response handler} stage. Like typical processing \stages{}, it uses multiple threads, with different priorities, to process responses from the primary and secondary replicas. Once the entire \emph{get()} operation is complete, the object/file is delivered to the application, and ongoing jobs at primary and secondary replicas corresponding to this \emph{get()} operation are purged. 

\paragraph{Support for HDFS.} HDFS does not queue requests natively. Instead of adding queues and other \stage{} primitives, we decided to test our Proxy \stage{} with unmodified HDFS datanodes. On the client side, we added support for dispatching -- creating and sending multiple requests to different replicas. Because HDFS uses a separate TCP connection for each request, we used the closure of a TCP connection as a purge signal as well.

\paragraph{Support for \hkust{}} 
We modified this application to introduce \stages{} at both the client and server. 
The bottleneck for the target workloads is the network, and the server already has support for adding prioritization to different network flows. 
The original server did not use queues, so we added support for queuing and purging. 
Given that typical target workloads for this application include small requests (e.g., few KBs), the system
uses persistent TCP connections, so we explicitly sent purge messages from client to server. 
These purge messages were sent using a separate TCP connection. Given the small request sizes, we only purged requests from the server queues as purging in-flight requests would not provide much savings. 
We also added support for pipelining of requests on the same TCP connection and used a fixed number of pre-established connections -- these optimizations
improved our baseline results (\textsf{Single-Copy}) compared to the vanilla implementation. 

\subsection{Case Study: NFV}
The case of NFV cluster is different from the data parallel applications in two ways: i) jobs are duplicated by an intermediate network \stage{} instead of by the application \stage{}, ii) purging is even more challenging because of extremely short timescale of (per packet) operations.

Similar to figure~\ref{fig:proc_pkts}, we consider a \snort{}~\cite{snort3} based IDS cluster deployment as a use case for \sched{}.
We have a network \stage{}, built on top of Packet Bricks~\cite{packet-bricks}, which is responsible for duplicating and load balancing incoming packets to IDS nodes.
On each node, a processing \stage{} runs primary and duplicate \snort{} instances on different threads pinned to the same core at high and low priority respectively. It uses the POSIX call $pthread\_setschedprio()$ to enable thread level CPU prioritization.
Finally, another network \stage{} sits after the IDS cluster and acts as a ``response handler''; 
it performs de-duplication and forwards only unique packets to an interested application (e.g., web server).
We approximate purging at processing \stages{} on IDS nodes by limiting the secondary \snort{}'s queue size.

\section{Implementation}
\label{sec:impl}
In addition to the above applications, we have also implemented a Proxy \stage{} in C++; it implements the basic \stage{} interface (with support for purging, prioritization) as well as the throttling based scheduler. The proxy is multi-threaded, supports TCP-based applications, with customized modules for the HDFS application, in order to understand its requests. 
The proxy has been evaluated with multiple applications for benchmarking purposes: it adds minimal overhead for applications with small workloads (e.g., \hkust{}) and for the applications where we actually use the Proxy (i.e., HDFS), the performance matches that of the baseline (without the proxy). 
The proxy also supports a module to directly interact with the storage, and a client interface that can be used to generate \emph{get()} requests (similar to the HDFS and \hkust{} applications).
For prioritization we use the different primitives available for each resource (as described earlier). Specifically, for the network, we used native support for priority queuing, including Linux HTB queues at end-points. For storage, Linux provides $ioprio\_set()$~\cite{manpage-ioprio-set} system calls to explicitly set I/O prioritization of a request.
\section{Evaluation}
\label{sec:eval}

Our evaluation covers a broad spectrum, in terms of environments (public cloud and controlled testbed), applications (HDFS, \hkust{}, \snort{} IDS), system load (low, medium, high\footnote{By our definition, low load is 10-20\%, medium load is 40-50\%, high load is 70-80\%}), workloads (with mean RCT from $\mu$s to ms and seconds), schemes against which \sched{} is evaluated (e.g., \textsf{Hedged}, \textsf{Cloning}, etc), and micro-benchmarking various aspects of our system (e.g., prioritization overhead, etc). Our key insights are:

\begin{itemize}[leftmargin=*,noitemsep,topsep=0ex]

\item \textbf{\sched{} effectively avoids different types of stragglers observed in the ``wild''.} In our experiments on the public cloud, we encounter different types of stragglers (e.g. caused by storage and network bottlenecks). \sched{} is able to avoid most of these stragglers resulting in up to a 4.6$\times$ reduction in tail latency (p99 RCT) compared to the baseline. (\S\ref{subsec:eval-hdfs-cloud}) 

\item \textbf{\sched{} is safe and easy to use.}
Using controlled experiments on Emulab, we show that \sched{} remains stable at high loads, and performs as well as the best performing duplication scheme
under various scenarios -- all these benefits come without requiring any fine tuning of thresholds. (\S\ref{subsec:eval-hdfs-emulab} and \S\ref{subsec:eval-network}) 

\item \textbf{\sched{} can effectively deal with system and workload heterogeneity.} A (somewhat) surprising finding of our study is that the use of replicas through
\sched{} can also shield small flows from large flows (workload heterogeneity) without requiring flow-size information (or approximation),  as is required by most datacenter transports~\cite{pias, pase}, also obviating the need for any intelligent replica selection mechanism~\cite{c3, sinbad}. (\S\ref{subsec:eval-network}) 

\item \textbf{\sched{} is feasible for an IDS cluster.}
We show that dealing with lower priority duplicate traffic does not affect the throughput of \snort{} nodes, and \sched{} is able to effectively deal with stragglers. (\S\ref{subsec:snort})

\end{itemize}

\subsection{HDFS Evaluation in Public Cloud Settings}
\label{subsec:eval-hdfs-cloud}

\begin{figure}[!t]
  \begin{center}
    \includegraphics[width=\columnwidth]{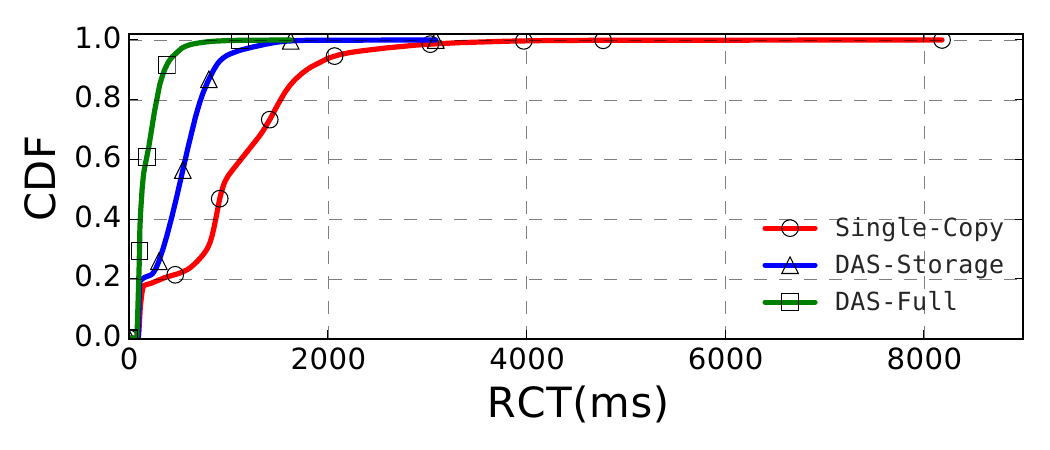}
  \end{center}
  \vspace{-0.2in}
  \caption{\sched{}'s performance in public cloud for HDFS cluster. \sched{} reduces latency at p99 by 4.6$\times$ and mean latency by 5.4$\times$.}
  \label{fig:GCP-40}
%   \vspace{-0.1in}
\end{figure}

\begin{figure}[!t]
  \begin{center}
    \includegraphics[width=\columnwidth]{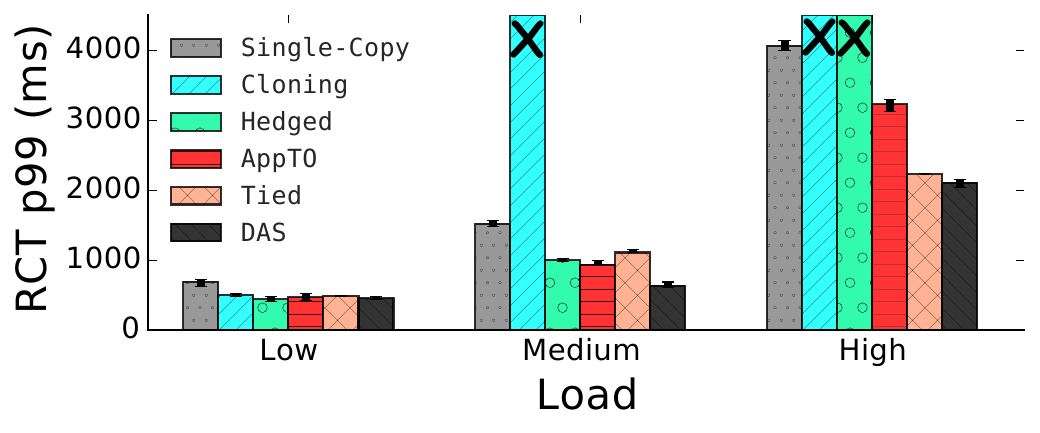}
  \end{center}
   \vspace{-0.2in}
  \caption{\sched{}'s comparison with other duplication based schemes. \sched{} performs equally well at low and medium loads while keeping the system stable at high load. Max-sized bar with black cross indicates that the corresponding scheme becomes unstable.}
  \label{fig:emulab-full-10MB}
%   \vspace{-0.1in}
\end{figure}

The goal of this experiment is to evaluate the robustness of our system in the ``wild'' -- this environment has natural stragglers and also involves multiple potential bottleneck resources (e.g., network, storage). 

\textbf{Experimental Setup.}
We set up an HDFS storage cluster on 70 VMs on Google Cloud~\cite{google-cloud}. Our cluster has 60 data-nodes and 10 clients. For HDFS data-nodes, we use \texttt{n1-standard-2} type machines (2 vCPUs, 7.5 GB RAM), while clients run on \texttt{n1-standard-4} machines (4 vCPUs, 15GB Memory). The HDFS replication factor is set to 3. We provision a total of 6TB of persistent storage\cite{GPC-persistent-disk} backed by standard hard disk drives. Persistent disk storage is not locally attached to the systems so the bottleneck could be anywhere in the system. On each data-node, we have two types of \stages{} that are used: a Proxy \stage{} for handling the HDFS data-node's unmodified storage \stage{}, and a network \stage{}, which makes the end-host support \sched{} while using the cloud network\footnote{Cloud networks present the big switch abstraction, with network bottlenecks only appearing at the edges.}.
 
\textbf{Workloads and Metric.}
For our experiments, the client generates \texttt{get()} requests for a fixed size file (10MB), according to a Poisson process. 
We run this experiment at medium load (estimated from the disk throughput guaranteed by the cloud provider which translates to more than 2000 requests per minute) for several hours.
Our selection of file size corroborates with the small file sizes\footnote{In their study of Facebook Messages stack, Harter et al.\cite{facebookMessages} revisited the assumption that HDFS file sizes are large and read accesses are sequential. They found that 90\% of files are smaller than 15MB and disk I/O is highly random.} observed in the HDFS deployment at Facebook for their messages stack~\cite{facebookMessages}. The size of our data set is $\sim$1.5TB. Our evaluation metric in all of our experiments is the request completion time (RCT) unless specified otherwise. 
Note that each job comprises of a single request -- with a higher scale out factor (multiple requests in a single job), we can expect even more stragglers to show up~\cite{mittos}.

\textbf{Schemes.}
For this experiment, we compare the performance of \textsf{Single-Copy} (base-case) with \sched{} under two different settings: i) \sched{}-Storage, which only uses the storage \stage{}, ii) \sched{}-Full which uses both storage and network \stages{}. %We measured performance of both these schemes during different hours of the day, alternatively running both these %schemes. 

\textbf{Avoiding Stragglers at Multiple Layers.} 
Figure \ref{fig:GCP-40} shows the CDF of RCTs for the three schemes. 
We observe a long tail for the baseline: at p99.9, the latency is ~5$\times$ higher than the median latency (3.5$\times$ and 2.3$\times$ at p99 and p95 respectively). 
The results show that \sched{} reduces the tail by 2.26$\times$ at p99 with only storage \stage{} enabled. However, with both the storage and the network \stages{} (\sched{}-Full) yields higher gains; RCTs are reduced at all percentiles with the benefits more pronounced at higher percentiles (p99 latency is reduced by 4.6$\times$). 

Our analysis of the results revealed that the long tail was primarily caused by two different sources of stragglers: i) poorly performing datanodes (at p99 only 6 datanodes contributed $\sim$40\% to the tail, while 9 datanodes contributed $\sim$54\%), ii) network noise, in \textsf{Single-Copy} case, even the best performing datanode (datanode that contributed least to the tail) had p99 latency 2.4$\times$ higher than p99 latency achieved by \sched{}. With only storage \stage{} enabled (\sched{}\textsf{-Storage-Only}) we were able to handle first type of stragglers. However, we still observed that the best performing replica had p99 latency $\sim$1.1$\times$ higher than overall p99 latency. By enabling network \stage{} on datanodes, we handled the second type of stragglers as well. The best performing datanode had p99 latency $\sim$15\% less than p99 latency achieved by \sched{}\textsf{-Full}. This highlights the ability of \sched{} in dealing with different sources of stragglers, which may appear at different layers of the system. 

\subsection{HDFS Evaluation in Controlled Settings}
\label{subsec:eval-hdfs-emulab}
We now move to the controlled environment of Emulab,
to evaluate \sched{} against other duplication schemes 
under various scenarios. 
This environment, combined with our use of a straggler/noise model
derived from our previous experiment, allow us to perform repeatable, yet realistic experiments. 

\noindent
\textbf{Schemes.}
We compare the performance of \sched{} against \textsf{Single-Copy} (base case), and several duplication schemes proposed by Dean \etal~\cite{tail_at_scale}: \textsf{Cloning}, \textsf{Hedged}, \textsf{AppTO}, and \textsf{Tied}~\cite{tail_at_scale}. 
For \textsf{Cloning}, the client proactively duplicates every request at the same priority. For \textsf{Hedged}, if the first request fails to finish before p95 deadline, the client makes a duplicate request to a different replica. Upon completion of any of the two copies, the other is purged. In \textsf{AppTO}, requests have a timeout value. If a request fails to finish before the timer expires, the client purges the first request and issues its duplicate request to another replica (request restart). In our experiments we use 480ms as the threshold for triggering a duplicate request for \textsf{Hedged} and \textsf{AppTO} schemes.
The \textsf{Tied} scheme duplicates every request and ties the identity of the other replica with the request.
When a request's turn arrives, it sends a purge message to its counterpart. The corresponding request, if not finished, gets purged.

\noindent
\textbf{Experimental Setup.}
For this experiment, we setup a 10 node HDFS  cluster on Emulab \cite{emulab}. We provision one client for this experiment. We use d430 type machine (2x2.4 GHz 8-core with 64GB RAM, 1TB 7200 RPM 6 Gbps SATA Disks). HDFS datanodes are connected by 1Gbps network link; whereas, client is connected by 10Gbps network link. For random read I/O requests we benchmark the disk throughput to be less than the network link capacity making disks the bottleneck resource. This experiments uses the same workload and metrics as in the previous experiment.

\noindent
\textbf{Noise Model.}
To model stragglers, we derive a noise model from our experiment on the Google Cloud. We use the median latency experienced by \textsf{Single-Copy} requests as our baseline and calculate the additional latency experienced by the top 10 percentiles (p91 to p100). We then model this noise for 10\% of the requests in our experiment by delaying the application response by the calculated factor.

\paragraph{Results.}
Figure~\ref{fig:emulab-full-10MB} compares the performance of all the schemes at the p99 latency across different loads. 
The load corresponds to three regimes: low, medium, and high. This is with respect to the offered load at the bottleneck (i.e., disk)\footnote{Note that this is the load by the primary requests only. Duplication may increase the load depending on the particular scheme}.
%There are three important observations: 
We make four observations:
\begin{enumerate}[leftmargin=*]
    \item At low loads, all the duplication based schemes perform well compared to \textsf{Single-Copy}. They reduce the p99 latency by \emph{at least} $\sim$1.45$\times$.

    \item At medium and high loads, the duplication overheads of aggressive schemes (\textsf{Cloning}, \textsf{Hedged}) make the system unstable. \sched{}, despite duplicating every request, remains stable and continues to reduce tail latency (p99 by 2.37$\times$, 1.9$\times$ at medium and high loads respectively).
    
    \item \sched{} is most effective at medium load. This is because at medium load, transient load imbalance on any one node is common despite the overall load being moderate. This enables \sched{} to leverage the spare capacity on other nodes. In contrast, at low load, stragglers are less common at low load while there are little opportunities to exploit an alternate replica at high loads.

    \item \textsf{Tied} is useful in terms of system stability (does not cause system overload at high load). 
    However, it fails to cope with noise encountered once the request has started being served, which is evident from low gains at low and medium loads. \sched{} successfully handles such scenarios by continuing to work on both requests until one finishes. Further, as we show in \S\ref{subsec:eval-network}, the lack of prioritization in \textsf{Tied} is catastrophic for workloads with small requests where \textsf{Tied} becomes unstable at high loads (see Fig.~\ref{fig:network-99}) --  in such scenarios, purging alone is insufficient. 
    
\end{enumerate}
\subsection{\hkust{} in Public Cloud Settings}
\label{subsec:eval-network}

\begin{figure*}[!t]
    \centering
      \subfigure[]{\includegraphics[width=2.20in]{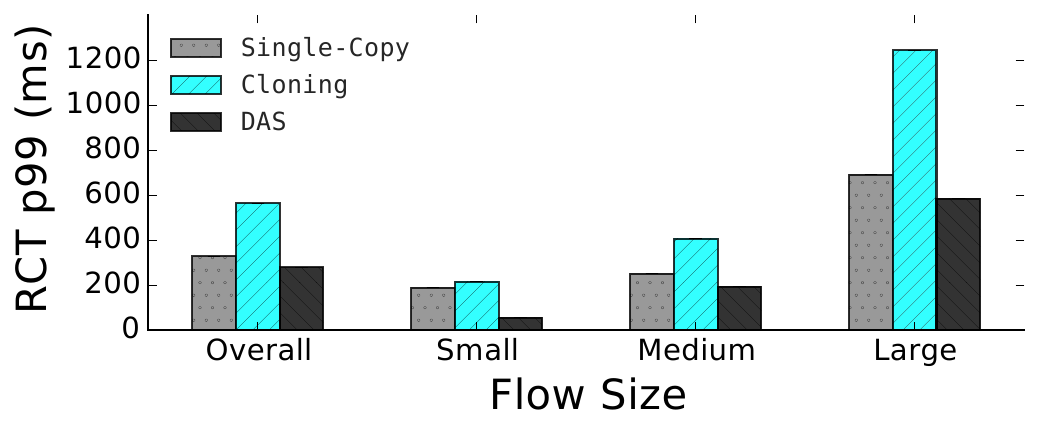}\label{fig:network_Mixed_99}}
      \subfigure[]{\includegraphics[width=2.20in]{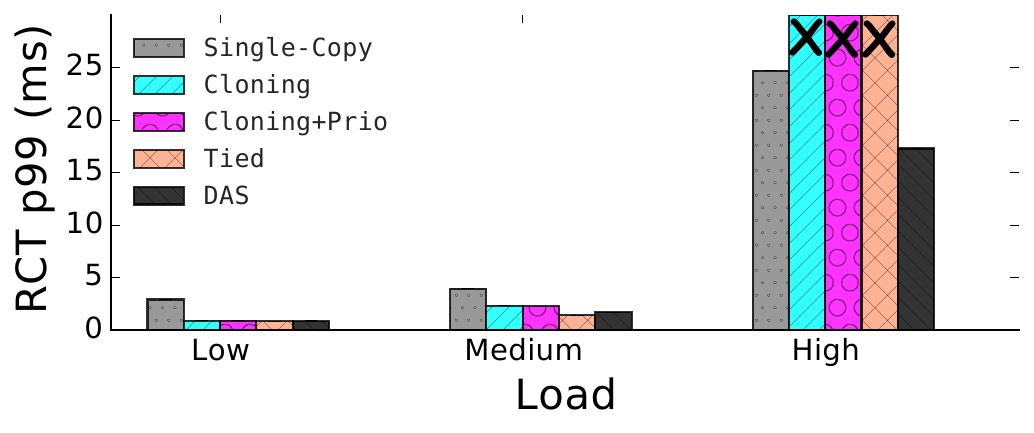}\label{fig:network-99}}
      \subfigure[]{\includegraphics[width=2.20in]{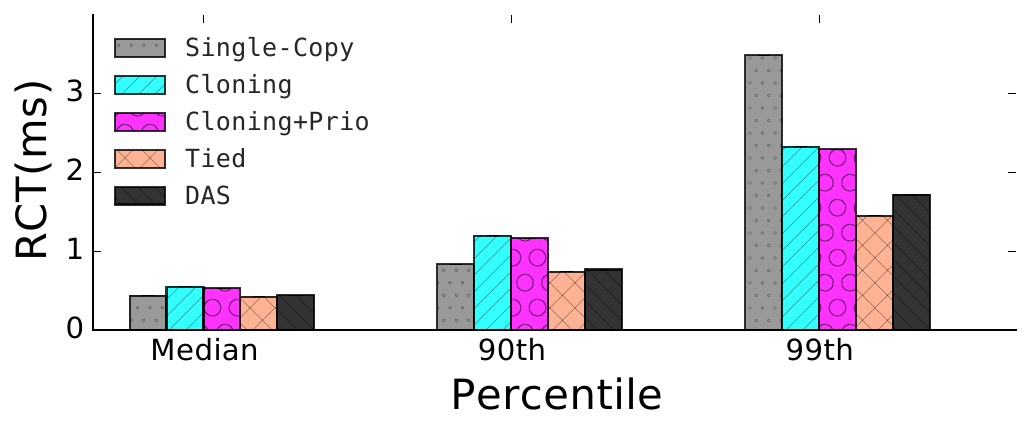}\label{fig:network-medium}}
      
      \caption{\sched{} under network bottleneck scenarios. (a) RCTs at p99 for web search workload~\cite{dctcp12} at medium load for different flow sizes -- Small (<100KB), Medium (100KB to 10MB), Large (>10MB) (b) RCTs at p99 across varying load when the flow sizes are fixed at 50KB. At high load, DAS keeps the system stable while other duplication based schemes fail (c) performance for fixed 50KB transfers at medium load at important percentiles. \sched{} is within 10\% of Tied Request while considerably better than other strategies at the tail ($\geq$ p90). Max-sized bar with black cross indicates that the corresponding scheme becomes unstable.}
\label{fig:network-bottleneck}
\end{figure*}

The goal of these experiments is to evaluate the performance under scenarios where the \emph{network} is the bottleneck. 
Compared to HDFS, the workload is also different (smaller requests) which creates new challenges and opportunities. 

\textbf{Setup.} We use a 10 VM setup on Google Cloud, with one client and nine servers. On the client side we use \texttt{n1-highcpu-16} type VM while servers use the \texttt{n1-stand-4} VM types. The server VMs are rate limited to 1Gbps while the client VMs have a limit of 16Gbps. 
We focus on small flows, using the average flow size (i.e., 50KB) of short flows from the web search workload~\cite{dctcp12} for most of our experiments.
We also consider the full web search workload, which includes large flows as well, and highlight how \sched{} is able to effectively deal with flow size heterogeneity. Our client generates requests based on a Poisson process, randomly choosing the primary and secondary servers.

\paragraph{Performance Under Workload Heterogeneity.}
Figure~\ref{fig:network_Mixed_99} shows the performance of \sched{} with the DCTCP web search workload (containing a mix of short and long flows) under medium load. \sched{} significantly improves the RCT for short flows without hurting the long ones. 
This happens because the co-existence of long and short flows significantly affects the RCTs of short flows~\cite{dctcp12}. 
\sched{} is a natural fit for such scenarios. If a short flow and a long flow get mapped to a critical resource together, namely a link or a queue, \sched{} provides opportunity to the duplicate request of short flows to get served from another less loaded replica. This highlights that if we have replicas and have a scheme like \sched{}, we can shield small flows from long flows without requiring flow size information (or estimation) as is required by most datacenter transports~\cite{pase}.
%\sched{} is a natural fit for such scenarios. If a short flow and a long flow get mapped to a critical resource together (e.g. a link), \sched{} provides opportunity to the duplicates of short flows to get served from less loaded replica. This highlights that if we have replicas and have a scheme like \sched{}, we can shield small flows from long flows without requiring flow size information (or estimation) as is required by most datacenter transports~\cite{pase}.

\paragraph{Performance with Small Requests.}
We now focus on fixed small (50KB) request sizes. 
We introduce noise through background flows: the client randomly picks a server and starts a 10MB transfer,
%emulating a network hotspot. Overall, the noise accounts for only 1\% of the total link capacity.
emulating a hotspot in the network. Overall, the noise accounts for only 1\% of the total link capacity.

Figure~\ref{fig:network-99} shows the performance of \sched{} at the p99 across variable network loads -- low, medium, and high. Similar to the HDFS experiments, we observe that: \sched{} remains stable at high loads while providing duplication gains at low loads. Note that no other duplication scheme remains stable at high loads; this is unlike the HDFS experiment (where schemes like Tied worked well), and is because of the challenging nature of this workload which involves small (sub-millisecond) requests. 
In such scenarios, purging is less effective at low loads (in fact, it has overhead) and prioritization becomes critical, as highlighted by the slight difference in the performance of \sched{} and the \textsf{Cloning+Prioritization} scheme. However, our analysis shows that at high loads we need \emph{both} prioritization and purging in order to keep the system stable.
Figure~\ref{fig:network-medium} zooms into the medium load and highlights achieved gains over \textsf{Cloning} (unstable at high load) and \textsf{Single-copy} (stable at high load) at different percentiles. It is also within 10\% of the \textsf{Tied} scheme which becomes unstable under high load.

\subsection{\sched{} with IDS cluster}
\label{subsec:snort}
We evaluate the feasibility of using \sched{} when CPU is the bottleneck in the context of an IDS cluster. This can be useful in scenarios where some nodes are stragglers while others have spare CPU capacity to process traffic.

We consider an IDS cluster receiving packets at varying load (defined as the arrival rate of traffic).
For each load, we determine the performance of having single snort (\textsf{Single-Copy}) instances versus having two snort (Primary and Duplicate) instances. This comparison highlights that duplication of work does not hurt performance. Secondly, for each load, we consider the effect of having a straggler node. This comparison highlights that duplication can alleviate the problem of stragglers.

\begin{figure}
    \centering
      \subfigure[Without straggler]{\includegraphics[width=0.47\columnwidth]{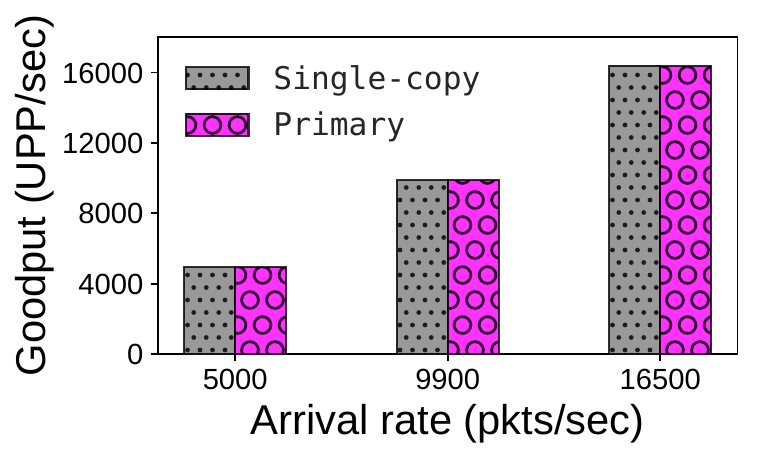}\label{fig:snort-a}}
      \subfigure[With straggler]{\includegraphics[width=0.47\columnwidth]{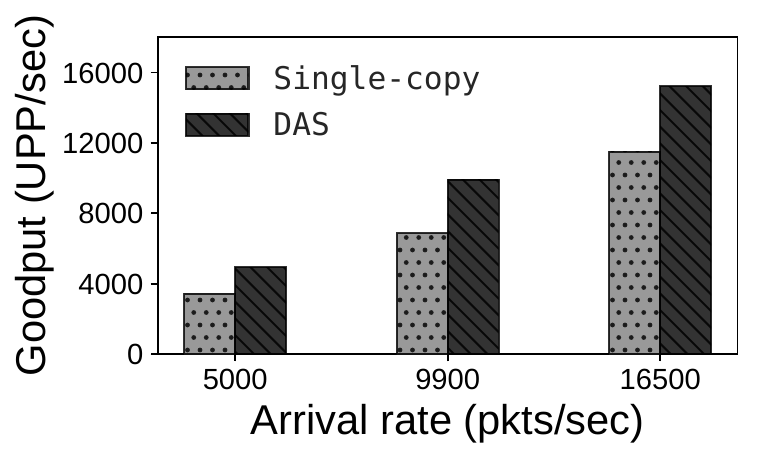}\label{fig:snort-b}}
      \caption{Feasibility of using duplication when CPU is the bottleneck in an IDS cluster. (a) shows that at varying input load, the introduction of a duplicate snort does not hurt primary snort throughput. (b) shows that with a straggler node, \sched{} outperforms \textsf{Single-Copy} as secondary snort instances running at other nodes can process the duplicate of the packets that are backlogged at the straggler.}
      %process the duplicates of the packets backlogged at the straggler.}
\label{fig:compute-bottleneck}
\end{figure}

\paragraph{Experimental Setup.} For this experiment we provision five d430 type machines on Emulab. Our setup reflects the layout given in Figure~\ref{fig:proc_pkts} (\S\ref{subsec:target-scen}). 
The \snort{} IDS runs on two nodes; each node runs a primary and duplicate instance of \snort{} at high and low priority respectively.

On each node, \snort{} instances are pinned to the same core, and we use a single thread for the \snort{} instance.
The other three nodes are configured as a traffic source, an in-network packet duplicator and an in-network de-duplicator.

A straggler in this setting could be due to system overload or some failure~\cite{bugs_in_cloud,bugs_in_cloud2}.
To emulate this behaviour, we run a CPU-intensive background task lasting 60\% of the experiment duration on one of the IDS nodes. 
This effectively results in a drop in \snort{}'s throughput.
We focus on the goodput -- unique packets processed per second (UPP/sec) -- achieved by the IDS cluster under varying system load.

\paragraph{Prioritization Works.} 
Figure~\ref{fig:snort-a} shows the goodput for varying system loads without any stragglers.
We observe that with thread prioritization the impact of duplicate \snort{} on the goodput of primary \snort{} is negligible -- even when the traffic arrival rate is high.

\paragraph{\sched{} Alleviates Stragglers.}
Figure~\ref{fig:snort-b} shows the performance of \snort{} for varying system load when one \snort{} node becomes a straggler.
We observe that goodput in the case of \sched{} is higher than \textsf{Single-Copy} (no duplication). This is because packets dropped by straggler node in the case of \textsf{Single-Copy} are processed by the duplicate \snort{} instance in the case of \sched{}. 
\subsection{Microbenchmark Results}
\label{subsec:eval-micro}

In this section we scrutinize the: 
i) overhead of disk I/O prioritization, and ii) overhead of network prioritization.
\begin{figure}
  \begin{center}
    \subfigure[Network Prioritization]{
        \includegraphics[width=0.47\columnwidth]{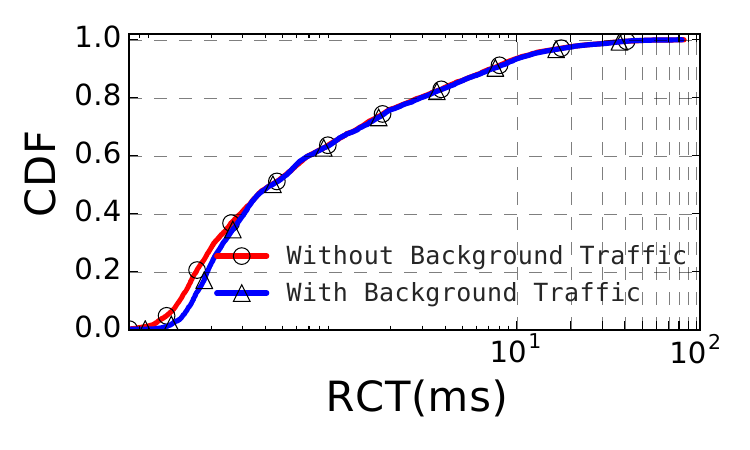}
        \label{fig:net-micro}}
    \subfigure[I/O Prioritization]{
        \includegraphics[width=0.47\columnwidth]{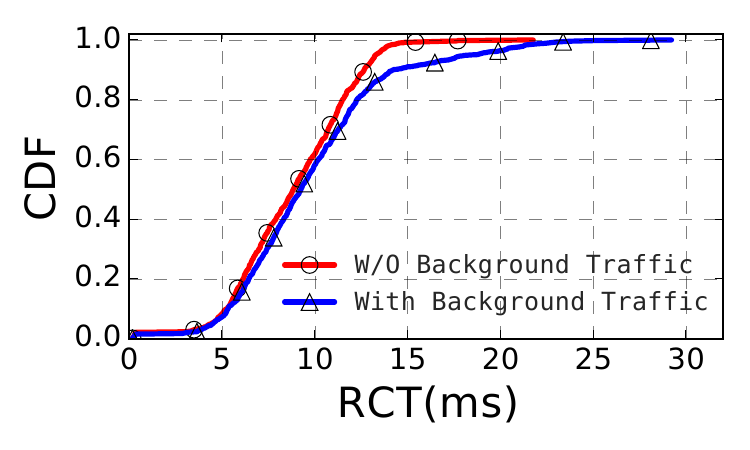}
        \label{fig:storage-micro}}
  \end{center}
  \caption{Prioritization benchmarks showing RCTs of high priority requests with and without low priority background traffic.} 
  \label{fig:microbenchmarking}
\end{figure}

 \paragraph{Efficiency of network prioritization.}
We investigate the overhead of the Linux network prioritization mechanism, incurred by high priority requests in the presence of a low priority network flow. 
We set up a client connected to a server on a 10Gbps link. The client runs the \hkust{} application and generates requests according to the web search workload (\cite{dctcp12}) at medium load. The ToS bit for all these requests are marked as ``high'' while a single long TCP flow runs in the background at low priority.
Figure~\ref{fig:net-micro} shows the CDF of the RCTs of the high priority requests. Ideally, adding background traffic shouldn't have an impact on the RCTs of the high priority requests due to the enforced prioritization. We observe an overhead of $\leq$ 20\% at p10, at p20 it is $\leq$ 15\% and at p40 and beyond it is $\leq$ 3\%. This shows that network prioritization has some overhead for small requests (under 20\%), but the gains 
due to duplication outweigh this overhead, as shown earlier in our macrobenchmark experiments.

\paragraph{Efficiency of disk I/O prioritization.}
We evaluate the efficiency of Linux's CFQ~\cite{kerneldoc-CFQ} scheduler which we used for disk I/O prioritization. 
For this experiment, we deploy a 100GB data set on a Linux server and run variable number of high priority and low priority workers. Each worker continuously requests random 4KB blocks in a closed loop.

Figure \ref{fig:storage-micro} shows the CDF of RCTs of high priority requests with and without low priority background requests. We observe that roughly 20\%  of these requests are hurt by the presence of low priority requests. 
While for realistic workloads the gains of duplication outweigh this penalty, we have also verified that this overhead can be further reduced by tuning CFQ configuration parameters (e.g., \emph{slice\_sync, slice\_idle\_time}, etc).
\section{Related Work}

\noindent \textbf{Other Duplication Policies.}
To supplement the earlier discussion in \S\ref{subsec:existing-schemes}, we comment 
on the most relevant duplication policies. 
RepFlow~\cite{repflow} replicates certain flows (but at the \emph{same} priority) while Vulimiri~\etal~\cite{lowlatency} conduct some preliminary experiments using redundant flows with lower priority but do not consider the full set of features we propose in \sched{} (e.g., purging) or address system design challenges. 
Other duplication schemes have highlighted the benefits of prioritization through analysis and 
simulations~\cite{gardnerthesis, GARDNER2017, rans-hotnets}. We show that for practical systems, in addition to prioritization, purging \emph{must} be used.
Further, none of the above schemes focus on an abstraction for supporting duplicates. 
A recent work, MittOS introduces a fast rejecting SLO-aware interface to support millisecond tail tolerance but only consider storage stacks ~\cite{mittos} -- its technique to reject a request based on expected latency needs is highly customized for storage.

\noindent \textbf{Stage Abstraction.}
Welsh et al. proposed the staged event-driven architecture (SEDA) for designing scalable Internet services~\cite{seda}.
In SEDA, applications consist of a network of event-driven stages connected by explicit queues. In contrast, we consider \emph{duplicate-aware} stages (\stages{}), where each stage may correspond to a different resources (e.g., network, compute, and storage).
IOFlow \cite{ioflow} introduces the abstraction of a data plane stage for storage systems but does not consider duplication. We show that duplication brings its own unique challenges in designing stages for each bottleneck resource and structuring their interactions.

\noindent \textbf{Resource Capacity Estimation.}
Our Proxy \stage{}'s use of capacity estimation in its throttling mechanism is similar to 
resource capacity estimation in systems like PARDA~\cite{parda} and VDC~\cite{vdc}. 
Similar to us, both these systems use capacity estimation techniques that are inspired by TCP. 
However, these systems focus on dealing with increased latency or SLO violations whereas our goal is to maintain maximal control at the Proxy \stage{} while avoiding under utilization.
\section{Discussion and Future Work} 
\label{sec:discuss}

\paragraph{Duplicates with User-level Network Stacks.}
The support for duplicates can be effectively introduced in today's high performance user-level network stacks~\cite{mtcp, fstack} that use kernel-bypass and leverage network I/O libraries such as DPDK~\cite{dpdk} or Netmap~\cite{netmap}.
Existing NICs provide support for multiple queues and by applying appropriate filters (e.g., by using Intel's Ethernet Flow Director), duplicate traffic can be pinned to separate queues. These queues may then be served by a (strictly) low priority thread.

\paragraph{Work Aggregation.} 
Making duplicates safe to use also opens up another opportunity: work aggregation. This requires the scheduler to do \emph{fine-grained work},  allowing \emph{aggregation} of the fine-grained work done by the the primary and duplicate copies of the job. For example, consider a job -- having two sub-parts (A and B) -- that is being processed at two different stages; the primary copy can first process part A while the duplicate copy processes part B -- aggregating these parts can allow the job to finish even though both the primary and duplicate copies are not individually finished. These incomplete copies can subsequently be purged from the system. This feature requires a complementary processing \stage{} where the fine-grained work can be aggregated.
Identifying layers where such aggregation may be feasible, dealing with the associated challenges and overhead of doing fine-grained work, and 
quantifying the benefits, will be the key questions to address for this approach.

\vspace{0.025in} \noindent \textbf{Redundancy-based Storage Systems.} 
Quorum-based and redundancy-based storage systems can also naturally deal with stragglers \cite{hitchhiker, trevi}, but they incur high reconstruction cost because they need to make more requests than required. Such systems can potentially use \stages{} to reduce overhead. For example, $K$ out of $N$ requests can be high priority while the other $N-K$ can be lower priority

\paragraph{Energy Considerations.} 
While making duplicate may increase the energy consumption, there are two factors which can limit this overhead or may even reduce the overall energy consumption: i) we use purging so the overall work done by the system (in terms of executing a request) may not necessarily be significant, and ii) the overall reduction in response time implies that requests stay in the system for less time, possibly consuming fewer system resources and thus lowering the energy consumption.

\section{Conclusion}
%Tail latency is a major problem for cloud applications. This paper showed that by making duplicate requests a first-class concept, we can proactively use replicas without worrying about overloading the system. 
Tail latency is a major problem for cloud applications. This paper showed that we can make the use of duplication as a first-class concept to mitigate stragglers without worrying about overloading the system.
To this end, we proposed a new scheduling technique (\sched{}), an accompanying abstraction (\stage{}) that helps realize \sched{} for different resources, and 
end-to-end case studies that highlight the feasibility and benefits of \sched{} for different bottleneck resources. 
While our evaluation shows the promise of this approach in dealing with stragglers under a wide range of realistic scenarios, our work also opens up several interesting avenues for future work, such as how to get duplication gains in scenarios with different bottlenecks. 

\paragraph{Acknowledgement.} This work was supported by NSF grant 1618321.

\setlength{\bibsep}{2pt plus 1pt}

%\bibliography{ref}
%\bibliographystyle{abbrvnat}

\end{document}